\documentclass[reprint,
groupedaddress,
showpacs,preprintnumbers,
 amsmath,amssymb,
 aps,
prl
]{revtex4-1}
\usepackage{graphicx}
\usepackage{dcolumn}
\usepackage{bm}
\usepackage{mathrsfs}

\begin{document}

\title{Tunneling spectroscopy and Josephson current of superconductor-ferromagnet hybrids on the surface of a 3D TI}

\author{Bo Lu$^{1,3}$, Pablo Burset$^{2}$, Keiji Yada$^{1}$, Yukio Tanaka$^{1,3}$}

\affiliation{$^1$~Department of Applied Physics, Nagoya University, Nagoya 464-8603, Japan\\
$^2$~Institute for Theoretical Physics and Astrophysics, University of W\"{u}rzburg, D-97074 W\"{u}rzburg, Germany\\
$^3$~Moscow Institute of Physics and Technology, Dolgoprudny, Moscow 141700, Russia
}
\date{\today}
\begin{abstract}
We investigate the charge transport property of superconductor (S) /normal metal (N) / ferromagnet insulator
(FI) /(normal metal) N' and S/N/FI/N'/S Josephson junctions on a three-dimensional topological insulator surface. We find the asymmetric local density of states (LDOSs) in a S/N/FI/N' junction and show that the N interlayer gives rise to subgap resonant spikes in the differential conductance and LDOSs. In a S/N/FI/N'/S junction, the Josephson current shows a non-sinusoidal current-phase relation and the N (or N') interlayer decreases the magnitude of the critical current monotonically.

\end{abstract}

\maketitle

\section{I. Introduction}

Three dimensional (3D) topological insulator (TI) is a phase of matter with topologically protected Dirac-type surface states on their time
reversal invariant point \cite{Fu07prb,Hsieh08,Hsieh09,Hasan09,ZSC09,Chen09,Xia09,Ando10,KaneReview,MooreReview,ZhangReview,AndoReview}. With coupling to a ferromagnet (F), Dirac fermions show many exotic
properties such as
magnetoelectric effect \cite{Qi08prb,Qi08np,Moore09,MacD10,Franz10,Nomura11}. By the proximity effect to a
superconductor (S), the 3D TI surface states may become a topological superconductor
\cite{Kane2008}. When F and S coexist on 3D TI surfaces, it
is found that chiral Majorana edge states can be generated at the boundary
between them \cite{Kane2008,Kane2009,Beenakker09}, which leads to the
formation of zero-biased conductance peak (ZBCP) \cite{TK95} as experimental signatures\cite%
{Tanaka2009,BeenakkerReview,AliceaReview,Tanakareview,Snelder15}. Intrinsic topological superconductivity has also been found in doped 3D TIs, e.g., Cu$_{x}$Bi$_{2}$Se$_{3}$ \cite{hor10,097001,Sasak11,Kriener11,levy13}.

On the other hand, a variety of interesting phenomena about Josephson effect
in TI materials have been discovered \cite%
{Zhang11,Sacepe,Brinkman,Williams,Wang12,Snelder14,Moler15}. Recently, a
non-sinusoidal current-phase relation has been reported in the 3D TI
HgTe junction \cite{Moler15}. In the 3D TI heterojunctions like Nb/Bi$_{1.5}$Sb$_{0.5}$Te$_{1.7}$Se$_{1.3}$/Nb, the
temperature dependence of the critical current is almost linear in most of
the range \cite{Snelder14}. Also, the novel Josephson effect involving
Majorana fermions has been predicted theoretically \cite%
{Tanaka2009,Linderprl10,Linderprb10,Yokoyama12,Snelder13}, however, there
has been no experimental report yet. The rapid development in experiments
requires for a theoretical approach which can deal with realistic structures for Josephson junctions on 3D TI surface.

In this article, we address how to compose Green's function by wave
functions on superconducting 3D TI surface. Using the resulting formalism, one can analyze
the spacial dependence of physical quantities, such as local density of
states (LDOSs) and pair potentials. Also, this approach provides an efficient
way to calculate Josephson current for realistic junctions on 3D TI surfaces. In this work, we
consider the S/normal metal (N)/ferromagnetic insulator (FI)/N' junction and S/N/FI/N'/S Josephson junction as examples. Since making direct contact between F and S regions is not easily accessible in actual experiments,
the presence of N interlayer between S and F is a more realistic
setup to study Majorana fermions. In the S/N/FI/N' junction, we find that
the conductance spectra and LDOSs have spikes as a function of bias voltage
and quasiparticle energy $E$, respectively. The resulting LDOSs shows an
asymmetric energy dependence around $E=0$ . For the S/N/FI/N'/S junction, we find that the distance of N
interlayer (or N' interlayer) decreases the critical current monotonically.
The junctions with or without FI show a non-sinusoidal
current-phase relation at low temperatures.

The paper is organized as follows: In section II, we introduce our model and construct the Green's function. In section III, we show numerical results for S/N/FI/N' and S/N/FI/N'/S junctions and discuss them. A conclusion remark is given in Section IV.

\section{II. Model}

\begin{figure}[tbph]
\begin{center}
\includegraphics[width = 66 mm]{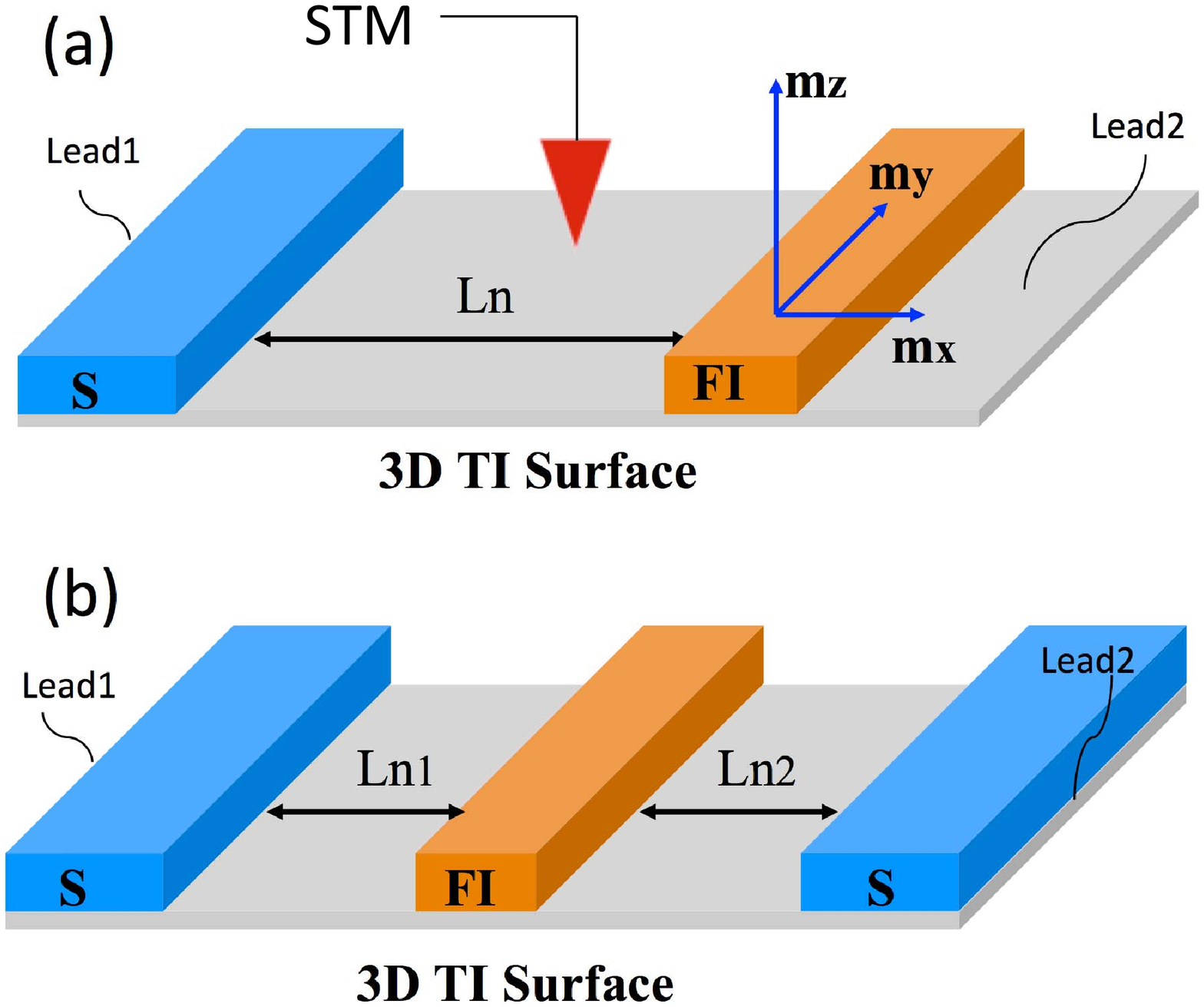}
\end{center}
\begin{center}
\caption{Schematics of the system: (a) S/N/FI/N' and (b) S/N/FI/N'/S formed on the surface of a 3D topological insulator. The local density of states can be detected by the STM tip. The differential conductance and the suppercurrent can be obtained from the leads on the two sides. }
\label{fig1}
\end{center}
\end{figure}

We consider the ballistic S/N/FI/N' and S/N/FI/N'/S junctions which are shown in Fig.%
\ref{fig1}.
The system can be described by the BdG Hamiltonian \cite%
{BdG,Tanaka2009}%
\begin{equation}
\hat{H}=\left[
\begin{array}{cc}
h(k_{x},k_{y})+M & i\hat{\sigma} _{y}\Delta  \\
-i\hat{\sigma} _{y}\Delta ^{\ast } & -h^{\ast }(-k_{x},-k_{y})-M^{\ast }%
\end{array}%
\right] ,  \label{Hamiltonian}
\end{equation}%
in $\left( \Psi _{\uparrow },\Psi _{\downarrow },\Psi _{\uparrow }^{\dag
},\Psi _{\downarrow }^{\dag }\right) ^{T}$ basis, where $%
h(k_{x},k_{y})=v_{f}(k_{y}\hat{\sigma} _{x}-k_{x}
\hat{\sigma} _{y})-\mu (\Theta \left(
-x+L_{n\left( n1\right) }\right) +\Theta \left( x-L_{n\left( n1\right)
}-L_{f}\right) )$  for the S/N/FI/N' (S/N/FI/N'/S) junction. $\hat{\sigma}%
_{i=x,y,z}$ are the Pauli matrices in the spin space and $\mu $ is the
chemical potential. Throughout the paper, we set $\hbar
=1$. The exchange field in F region is $M=\sum_{i=x,y,z}m_{i}%
\hat{\sigma}_{i}\Theta \left( x-L_{n\left( n1\right) }\right) \Theta \left(
L_{n\left( n1\right) }+L_{f}-x\right)$ for the S/N/FI/N' (S/N/FI/N'/S) junction. The pair potential $\Delta $ is given by $\Delta _{0}\Theta (-x)$ for
the S/N/FI/N' junction and $\Delta _{0}[\Theta (-x)+e^{-i\phi }\Theta
(x-L_{n1}-L_{f}-L_{n2})]$ for the S/N/FI/N'/S junction, where $\phi $ is the
macroscopic superconducting phase.

In this article, we use a standard formula of tunneling spectroscopy \cite%
{BTK,TK95} as shown in Ref.\cite{Tanaka2009} to obtain differential
conductance spectra of the S/N/FI/N' junction. Here, we would like to
present the way of constructing the retarded Green's function which has
recently been applied to relativistic system like Graphene \cite%
{Asano08,Burst10}, and 1D helical states on TI \cite{Pablo15}. In our system,
the translational invariance along the $y$-axis is preserved, thus the
retarded Green's function with respect to Eq.\ref{Hamiltonian} has the form $\check{G}(x,x^{\prime },y,y^{\prime
})=\sum\nolimits_{k_{y}}G^{k_{y}}(x,x^{\prime })e^{ik_{y}(y-y^{\prime })}$. The retarded Green's function can be written
as \cite{McMillan,FT,TanakaD,Tanaka2000,Burst10}%
\begin{equation}
\begin{array}{r}
G^{k_{y}}(x,x^{\prime })=\alpha _{1}\psi _{1}(x)\tilde{\psi}%
_{3}^{T}(x^{\prime })+\alpha _{2}\psi _{1}(x)\tilde{\psi}_{4}^{T}(x^{\prime
}) \\
+\alpha _{3}\psi _{2}(x)\tilde{\psi}_{3}^{T}(x^{\prime })+\alpha _{4}\psi
_{2}(x)\tilde{\psi}_{4}^{T}(x^{\prime }),%
\end{array}%
\end{equation}%
for $x>x^{\prime }$ and%
\begin{equation}
\begin{array}{r}
G^{k_{y}}(x,x^{\prime })=\beta _{1}\psi _{3}(x)\tilde{\psi}%
_{1}^{T}(x^{\prime })+\beta _{2}\psi _{4}(x)\tilde{\psi}_{1}^{T}(x^{\prime })
\\
+\beta _{3}\psi _{3}(x)\tilde{\psi}_{2}^{T}(x^{\prime })+\beta _{4}\psi
_{4}(x)\tilde{\psi}_{2}^{T}(x^{\prime }),%
\end{array}%
\end{equation}%
for $x<x^{\prime }$. $\psi _{i=1\sim 4}\left( x\right) $ are wave functions
of Eq.(\ref{Hamiltonian}) with wave vector $k_{y}$. $\psi _{1(2)}(x)$ is the
wave function for an incident electron-like (hole-like) particle from the
left side. $\psi _{3(4)}(x)$ is the wave function for the incident
electron-like (hole-like) particles from the right side. $\tilde{\psi}%
_{i=1\sim 4}(x^{\prime })$ are the wave functions corresponding to the
conjugate processes under the Hamiltonian
\begin{equation}
\tilde{H}=\left[
\begin{array}{cc}
\tilde{h}(k_{x},k_{y})+M^{\ast } & i\sigma _{y}\Delta ^{\ast } \\
-i\sigma _{y}\Delta & -\tilde{h}^{\ast }(-k_{x},-k_{y})-M%
\end{array}%
\right]  \label{cHamiltonian}
\end{equation}
with wave vector $-k_{y}$ and $\tilde{h}(k_{x},k_{y})$ is given by $\tilde{h}%
(k_{x},k_{y})=v_{f}(-k_{y}\sigma _{x}-k_{x}\sigma _{y})-\mu[\Theta
(-x+d_{n1})+\Theta (x-d_{n1}-d_{f})]$. For example, in the left S side, the
wave functions are
\begin{subequations}
\begin{alignat}{4}
& \psi _{1}(x)=\hat{A}_{1}e^{ik_{+}x}+a_{1}\hat{A}_{4}e^{ik_{-}x}+b_{1}\hat{A%
}_{3}e^{-ik_{+}x}, & & & & & & \\
& \psi _{2}(x)=\hat{A}_{2}e^{-ik_{-}x}+a_{2}\hat{A}_{3}e^{-ik_{+}x}+b_{2}%
\hat{A}_{4}e^{ik_{-}x}, & & & & & & \\
& \psi _{3}(x)=c_{3}\hat{A}_{3}e^{-ik_{+}x}+d_{3}\hat{A}_{4}e^{ik_{-}x}, & &
& & & & \\
& \psi _{4}(x)=c_{4}\hat{A}_{4}e^{ik_{-}x}+d_{4}\hat{A}_{3}e^{-ik_{+}x}, & &
& & & &
\end{alignat}%
and
\end{subequations}
\begin{subequations}
\begin{alignat}{4}
& \tilde{\psi}_{1}(x^{\prime })=\hat{B}_{1}e^{ik_{+}x^{\prime }}+\tilde{a}%
_{1}\hat{B}_{4}e^{ik_{-}x^{\prime }}+\tilde{b}_{1}\hat{B}_{3}e^{-ik_{+}x^{%
\prime }},\newline
& & & & & & \\
& \tilde{\psi}_{2}(x^{\prime })=\hat{B}_{2}e^{-ik_{-}x^{\prime }}+\tilde{a}%
_{2}\hat{B}_{3}e^{-ik_{+}x^{\prime }}+\tilde{b}_{2}\hat{B}%
_{4}e^{ik_{-}x^{\prime }},\newline
& & & & & & \\
& \tilde{\psi}_{3}(x^{\prime })=\tilde{c}_{3}\hat{B}_{3}e^{-ik_{+}x^{\prime
}}+\tilde{d}_{3}\hat{B}_{4}e^{ik_{-}x^{\prime }},\newline
& & & & & & \\
& \tilde{\psi}_{4}(x^{\prime })=\tilde{c}_{4}\hat{B}_{4}e^{ik_{-}x^{\prime
}}+\tilde{d}_{4}\hat{B}_{3}e^{-ik_{+}x^{\prime }}. & & & & & &
\end{alignat}%
The corresponding wave vectors are represented by $k_{\pm }=\sqrt{(\mu \pm
\sqrt{E^{2}-\Delta _{0}^{2}})^{2}/v_{f}^{2}-k_{y}^{2}}\equiv q_{e(h)}\cos
\theta _{\pm }$ and $q_{e(h)}=(\mu \pm \sqrt{E^{2}-\Delta _{0}^{2}})/v_{f}$.
The spinors are given as
\end{subequations}
\begin{subequations}
\begin{alignat}{4}
& \hat{A}_{1}(\hat{B}_{3}) & =& [iu,\pm e^{\pm i\theta _{+}}u,\mp e^{\pm
i\theta _{+}}v,iv]^{T}, & & & & \\
& \hat{A}_{2}(\hat{B}_{4}) & =& [ie^{\pm i\theta _{-}}v,\mp v,\pm u,ie^{\pm
i\theta _{-}}u]^{T}, & & & & \\
& \hat{A}_{3}(\hat{B}_{1}) & =& [ie^{\pm i\theta _{+}}u,\mp u,\pm v,ie^{\pm
i\theta _{+}}v]^{T}, & & & & \\
& \hat{A}_{4}(\hat{B}_{2}) & =& [iv,\pm e^{\pm i\theta _{-}}v,\mp e^{\pm
i\theta _{-}}u,iu]^{T}, & & & &
\end{alignat}%
where $u$ and $v$ are given by $u(v)$ $=$ $\sqrt{(E \pm\sqrt{E^{2}-\Delta_{0}
^{2}})/2E}$. Other wave functions can be found in the Appendix. The coefficients $a_{i}$, $b_{i}$, $\tilde{a}_{i}$ and $\tilde{b%
}_{i}$ can be solved from the boundary condition for relativistic
systems. For example, in S/N/FI/N' junction, the boundary conditions are: $\psi
_{i}(x=0_{+})=\psi _{i}(x=0_{-})$, $\psi _{i}(x=d_{n+})=\psi _{i,}(x=d_{n-})$%
, $\psi _{i}(x=d_{n}+d_{f+})=\psi _{i}(x=d_{n}+d_{f-})$, and similar to
other processes. $\alpha _{i=1\sim 4}$ and $\beta _{i=1\sim 4}$ can be
determined by the boundary conditions of Green's function
\end{subequations}
\begin{equation}
G^{k_{y}}(x+0,x)-G^{k_{y}}(x-0,x)=v_{f}^{-1}(i\hat{\tau}_{z}\hat{\sigma}%
_{y}),
\end{equation}%
where $\hat{\tau}_{i=x,y,z}$ are the Pauli matrices in the electron-hole
space. In real materials, the magnitude of the superconducting gap is much
smaller than the chemical potential $\Delta _{0}\sim 10^{-3}\mu $, so we can
use the quasiclassical approximation as $q_{e}\sim q_{h}$ and $\theta
_{+}\sim \theta _{-}\equiv \theta $. Then one can easily obtain the values
of $\alpha _{i=1\sim 4}$ and $\beta _{i=1\sim 4}$,
\begin{subequations}
\begin{eqnarray}
\alpha _{1(4)} &=&[2iv_{f}\cos \theta (u^{2}-v^{2})(\tilde{d}_{3}\tilde{d}%
_{4}-\tilde{c}_{3}\tilde{c}_{4})]^{-1}\tilde{c}_{4(3)}, \\
\alpha _{2(3)} &=&[2iv_{f}\cos \theta (u^{2}-v^{2})(\tilde{c}_{3}\tilde{c}%
_{4}-\tilde{d}_{3}\tilde{d}_{4})]^{-1}\tilde{d}_{3(4)}, \\
\beta _{1(4)} &=&[2iv_{f}\cos \theta
(u^{2}-v^{2})(d_{3}d_{4}-c_{3}c_{4})]^{-1}c_{4(3)}, \\
\beta _{2(3)} &=&[2iv_{f}\cos \theta
(u^{2}-v^{2})(c_{3}c_{4}-d_{3}d_{4})]^{-1}d_{3(4)}.
\end{eqnarray}%
From the Green's function, we can obtain the local density of states for
electrons: $\rho _{e}(x,E)$ and that for holes: $\rho _{h}(x,E)$,
\end{subequations}
\begin{equation}
\rho _{e\left( h\right) }(x,E)=\rho _{e\left( h\right) ,\uparrow }(x,E)+\rho
_{e\left( h\right) ,\downarrow }(x,E),
\end{equation}
where the spin-resolved LDOSs are given by%
\begin{eqnarray}
\rho _{e,\uparrow \left( \downarrow \right) }(x,E) &=&-\frac{1}{\pi }%
\sum_{k_{y}}\mathrm{Im}[G_{11\left( 22\right) }^{k_{y}}(x,x,E)], \\
\rho _{h,\uparrow \left( \downarrow \right) }(x,E) &=&-\frac{1}{\pi }%
\sum_{k_{y}}\mathrm{Im}[G_{33\left( 44\right) }^{k_{y}}(x,x,E)].
\end{eqnarray}
%
%
%
%
%
%
%
The dc Josephson current is determined by electric charge conservation rule
\begin{equation}
\partial _{t}P+\partial _{x}J_{x}+S=0,
\end{equation}%
where $P=\Psi _{\uparrow }^{\dag }\Psi _{\uparrow }+\Psi _{\downarrow
}^{\dag }\Psi _{\downarrow }$, $J_{x}=iv_{f}(\Psi _{\uparrow }^{\dag }\Psi
_{\downarrow }-\Psi _{\downarrow }^{\dag }\Psi _{\uparrow })$ and $S=2%
\mathrm{Im}[\Delta ^{\ast }\Psi _{\downarrow }\Psi _{\uparrow }-\Delta
^{\ast }\Psi _{\uparrow }\Psi _{\downarrow }]$ are electric charge density,
electric current and source term, respectively. After straightforward
calculations following Ref.\cite{FT,Tanaka2000}, we find that the total Josephson
current is
\begin{equation}
J_{x}=ek_{B}T\sum_{k_{y},\omega_{n}}\frac{\Delta}{2}\frac{\mathrm{sgn}%
(\omega_{n})}{\sqrt{\omega_{n}^{2}+\Delta ^{2}}}\left[a_{1}(i\omega
_{n})-a_{2}(i\omega_{n})\right],  \label{ft}
\end{equation}%
where $\omega _{n}$ is the Matsubara frequency $\omega _{n}=\pi
k_{B}T(2n+1),(n=0,\pm 1,\pm 2....)$. Eq.(\ref{ft}) shows that
Furusaki-Tsukada's formula \cite{FT} can also be applicable to the ballistic
Dirac-like electron systems on 3D TI surfaces \cite{Benj,bYang}. It enables us to
directly calculate the dc Josephson current in even more complicated or
long Josephson junctions on 3D TI surface without starting from the
energy levels of Andreev bound states \cite{Tanaka2009,Tkachov}.

\section{III. Numerical Results}
\subsection{A. S/N/FI/N' junction}
First, we show the conductance $\sigma _{s}$ (see Appendix) of S/N/FI/N' junction
in Fig.\ref{fig2}.
We normalized $\sigma _{s}$ by $\sigma _{n}$
which is the conductance when S is in normal
state. We only consider the exchange field along $z-$ and $x-$axis since the magnetization along $y-$axis does not change the
conductance \cite{Tanaka2009}. The length of the N layer between S and FI is denoted by $L_{n}$. The direct contact between S and FI means $L_{n}=0$.
For sufficient large $m_{z}(m_{x})$, the normalized conductance has a ZBCP similar to that in
chiral $p$-wave superconductor \cite{Tanaka2009} when magnetic field is along
$z$-axis as shown in Fig.\ref{fig2}(a).
Also we can see from Fig.\ref{fig2}(b), ZBCP appears when
the magnetization is along  $x-$axis. As $L_{n}$ increases, the sub-gap resonant peaks show up (Figs.\ref{fig2}(c)$\sim$(f)). The number of such peaks grows with $L_{n}$.
\begin{figure}[tbph]
\begin{center}
\includegraphics[width = 76 mm]{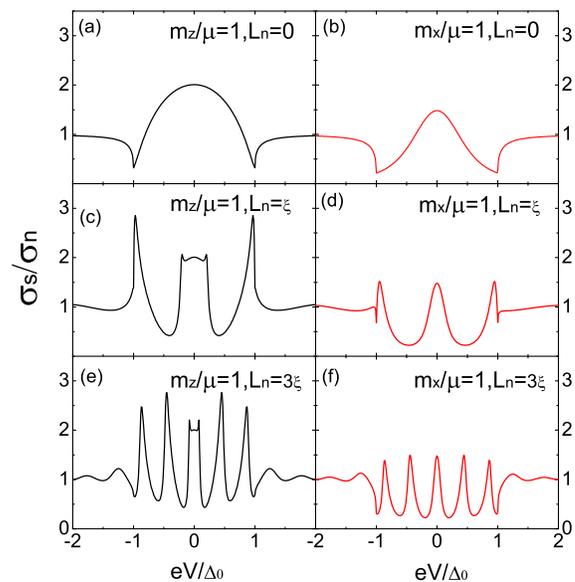}
\end{center}
\begin{center}
\caption{Normalized tunneling conductance as a function of bias voltage $(eV/\Delta_{0})$ for S/N/FI/N' junctions. (a), (b): $L_{n}=0$, (c), (d): $L_{n}=\xi$ and (e), (f): $L_{n}=3\xi$.
Black curve: $m_{z}/\mu=1$ and red curve: $m_{x}/\mu=1$. $\mu_=1$, $v_{f}$=1, $\Delta_{0}=0.001$ and $L_{f}=0.001\xi$ are chosen for all the panels.}
\label{fig2}
\end{center}
\end{figure}

\begin{figure}[tbph]
\begin{center}
\includegraphics[width = 76 mm]{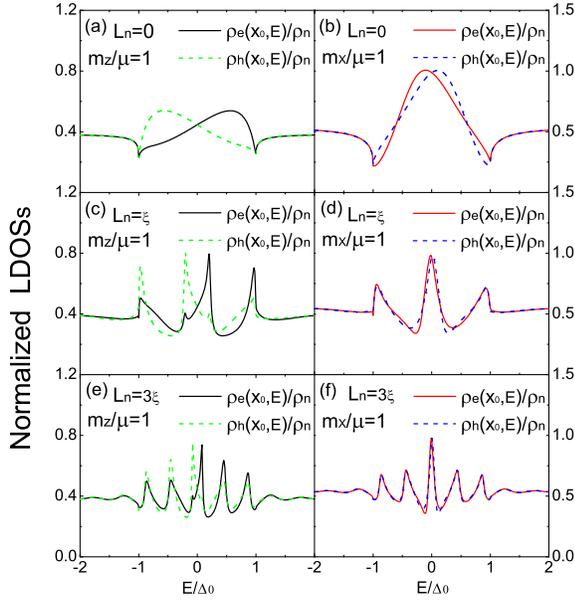}
\end{center}
\begin{center}
\caption{Local density of states in the middle of FI in the S/N/FI/N' junctions as a function of energy $(E/\Delta_{0})$: (a), (b): $L_{n}=0$, (c), (d): $L_{n}=\xi$ and (e),(f):$L_{n}=3\xi$.
Solid line for electron density of states and dashed line for hole density of states. Other parameters are chosen as the same as in Fig.\ref{fig2}. }
\label{fig3}
\end{center}
\end{figure}

This oscillatory phenomenon can also be seen in the local density of states $%
\rho _{e(h)}\left( x,E\right) $. We normalize $\rho _{e(h)}\left( x,E\right)
$ to that of the electron density of states of the bulk normal metal $\rho
_{n}$ at Fermi energy. Here, we choose the position in the middle of FI $%
x_{0}=L_{f}/2+L_{n}$ and show the density of states in Fig.\ref{fig3}.
When $L_{n}\neq 0$, we obtain the subgap peaks again as shown in Fig.\ref{fig3}(c $%
\sim $ f). The formation of such peaks can be explained as follows. We know
that the wave vector for electron (hole) is $k_{n}^{\pm}=\sqrt{(\mu \pm E)^{2}/v_{f}^{2}-k_{y}^{2}}$. The condition of forming the Andreev bound states in the N layer can be estimated from the Bohr-Sommerfeld quantization condition as
\begin{equation}
e^{i(k_{n}^{+}-k_{n}^{-})L_{n}}=1,
\end{equation}%
which shows that the number of peaks is proportional to $L_{n}$. Similar
formation of Andreev bound states was also revealed in junctions with 1D
helical edge states\cite{Crepin}.

We also find the asymmetric $E$ dependence of LDOSs near the S/FI interface,
e.g., $\rho _{e}\left( x_{0},E\right) $($\rho _{h}\left( x_{0},E\right) $)
in Fig.\ref{fig3}.
The asymmetry becomes prominent when magnetization is along $z$
axis (Figs.\ref{fig3}(a), (c) and (e)). We know that $\rho _{e}\left( x,E\right) $ and $\rho _{h}\left( x,E\right) $
are symmetric functions of $E$ for chiral $p$-wave superconductor when $\Delta_{0}$ is much smaller than $\mu$ \cite{Sigrist}. In that
case, the time-reversal symmetry is already broken in the bulk states of $p$%
-wave superconductor. On the other hand, the superconductor on TI is
time-reversal invariant and can not support chiral edge mode without
attaching ferromagnet. Therefore, we can imagine that
the chiral edge mode studied here has a nature similar
to  Shiba-type bound states \cite{Yu,Shiba,Rusinov}
by magnetic impurity scattering.

In usual case, where the spin degree of freedom is degenerate, the emerging
Shiba-states still follow the relation $\rho _{e\left( h\right) }(x,E)=\rho
_{e\left( h\right) }(x,-E)$, although the decomposed LDOS in each spin
sector $\rho _{e\left( h\right) ,\sigma }(x,E)$ does not satisfy $\rho
_{e,\sigma }(x,E)=\rho _{e,\sigma }(x,-E)$. Since $\rho _{e\left( h\right)
,\sigma }(x,E)=\rho _{e\left( h\right) ,-\sigma }(x,-E)$ is satisfied, after
summing up each spin component, $\rho _{e\left( h\right) }(x,E)=\rho
_{e\left( h\right) ,\uparrow }(x,E)+\rho _{e\left( h\right) ,\downarrow
}(x,E)=\rho _{e\left( h\right) ,\downarrow }(x,-E)+\rho _{e\left( h\right)
,\uparrow }(x,-E)=\rho _{e\left( h\right) }(x,-E)$ is satisfied. Then, the
resulting LDOS is symmetric around $E=0$. On the other hand, if the spin
degeneracy is lifted in the superconductor, it is possible that the LDOS
becomes asymmetric. In the present case, there is a strong spin-momentum
locking in the superconducting region by spin-orbit coupling. Then the
asymmetric energy dependence of $\rho _{e\left( h\right) }(x,E)$ appears
near the S/FI interface. In recent experiment of scanning tunneling
spectroscopy (STS), similar asymmetric behavior of LDOSs has been observed
in 1D S/F system\cite{Perge2014}. We can regard our finding in Fig. \ref{fig3} as
another example of asymmetric LDOSs in planar S/F junction which can be
detected in STS.

\begin{figure}[tbph]
\begin{center}
\includegraphics[width = 58 mm]{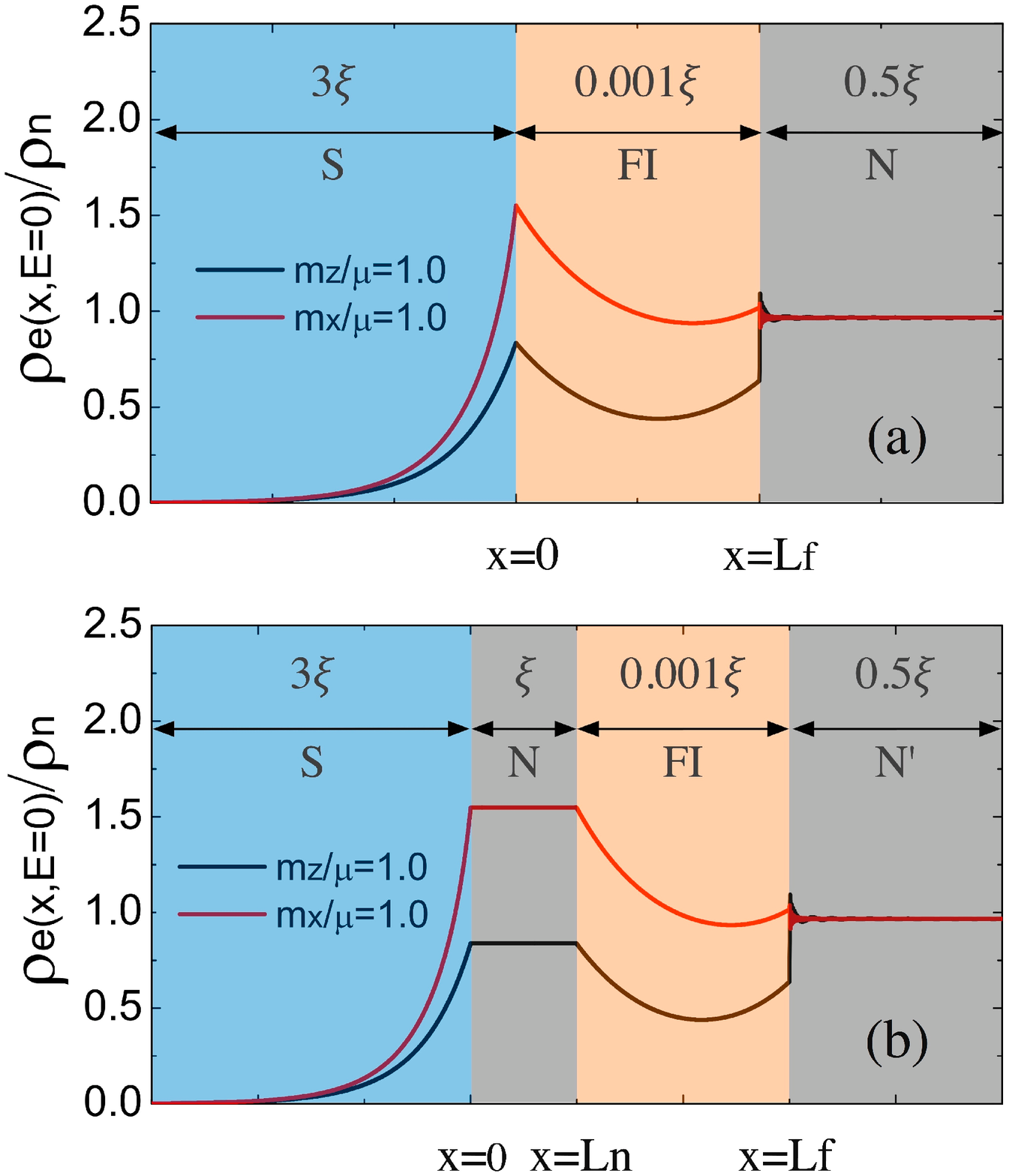}
\end{center}
\begin{center}
\caption{Spacial dependence of zero energy states in (a) S/FI/N junction and (b) S/N/FI/N' junction. The width of F layer is $L_{f}=0.001\xi$ and that of the N layer in (b) is $L_{n}=\xi$. Other parameters are the same as in Fig.\ref{fig2}. The scale of the horizontal axis is different in each region. }
\label{fig4}
\end{center}
\end{figure}

To see the spacial dependence of the Majorana states in such junctions, we
show the zero energy density of states $\rho _{e}(x,E=0)$ throughout the
junction. Because $\rho _{e}(x,E=0)$ is $0$ in both isolated S and FI
region, significant enhancement of $\rho _{e}(x,E=0)$ in S/FI interface of
S/FI/N junction can be regarded  as the experimental signature of chiral
Majorana fermion. In the S region, we can estimate that the characteristic length
expressing the spatial change of $\rho _{e}(x,E=0)$ is the order of macroscopic length scale: $\xi $.
This means a sufficient possibility to detect the presence of Majorana
fermion experimentally by STS, since the
manipulation of tip of STS just on the the S/N or S/F boundary with high
resolution is not easy.
Also, as seen in Fig.\ref{fig4}(b), even if there is a normal layer between
S and FI, the enhancement of $\rho _{e}(x,E=0)$ in both F and S is not
affected. In the N layer between S and FI, $\rho _{e}(x,E=0)$ is almost
constant. In the right N layer, we find oscillations of $\rho _{e}(x,E=0)$
on the scale of the inverse Fermi momenta. However, this oscillatory
behavior may be difficult to be detected in actual experiment.

\subsection{B. Josephson effect}
\begin{figure}[tbph]
\begin{center}
\includegraphics[width = 73 mm]{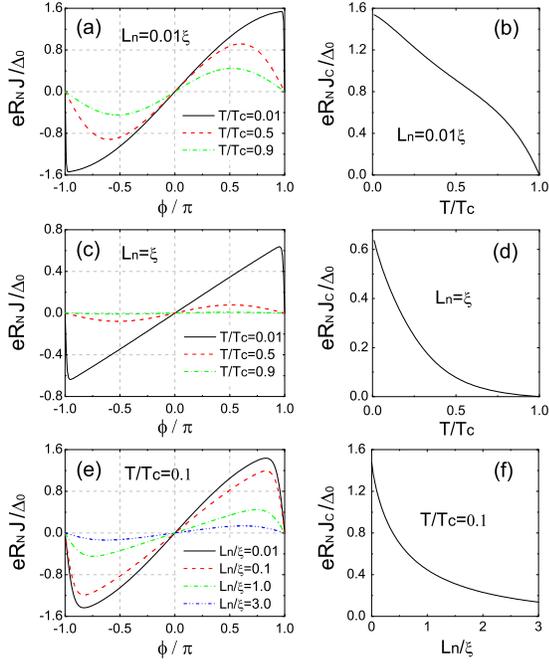}
\end{center}
\begin{center}
\caption{S/N/S Josephson junction: (a) Current-phase relation and (b) critical current for $L_{n}=0.01\xi$. (c) and (d) are those for $L_{n}=\xi$. Length dependence of $L_{n}$ for $T=0.1T_{c}$: (e) current-phase relation and (f) critical current. Other parameters are chosen as the same as in Fig.\ref{fig2}. }
\label{fig5}
\end{center}
\end{figure}
Before discussing the S/N/FI/N'/S junction, let us first look
at S/N/S junction.
Using Eq.(\ref{ft}), we plot the dc Josephson current in Fig.\ref{fig5}. It is normalized to $eR_{N}J/\Delta_{0}$ where $R_{N}$ is the interface resistance per unit area in the normal state. In panel (a), we can see that the current-phase relation is non-sinusoidal for short-junction in low temperature. This characteristic remains in the long-junction, as shown in panel (c). We notice that in recent experiment of Nb/3D-HgTe/Nb Josephson junctions, the current-phase relation is found to be non-sinusoidal \cite{Moler15}.
The experimental condition corresponds to
low temperature and long-junction in our calculation. We find a similar result in that limit as shown in (c).  The temperature dependence of critical current $J_{c}$ for short and long circumstances are given in panel (b) and (d), respectively. We observe that for high temperature, $J_{c}$ is a concave function of $T$ at small $L_{n}$ while it becomes a convex function with large $L_{n}$. It is also interesting to notice that in the large area of low temperature, $J_{c}$ is nearly a linear function of $T$ in both short- and long-junction. This result is in good agreement with the recent experiments in long Nb/Bi$_{1.5}$Sb$_{0.5}$Te$_{1.7}$Se$_{1.3}$/Nb Josephson junction \cite{Snelder14}. In Figs.\ref{fig5}(e) and (f), we plot the length dependence of Josephson current.

\begin{figure}[tbph]
\begin{center}
\includegraphics[width = 86 mm]{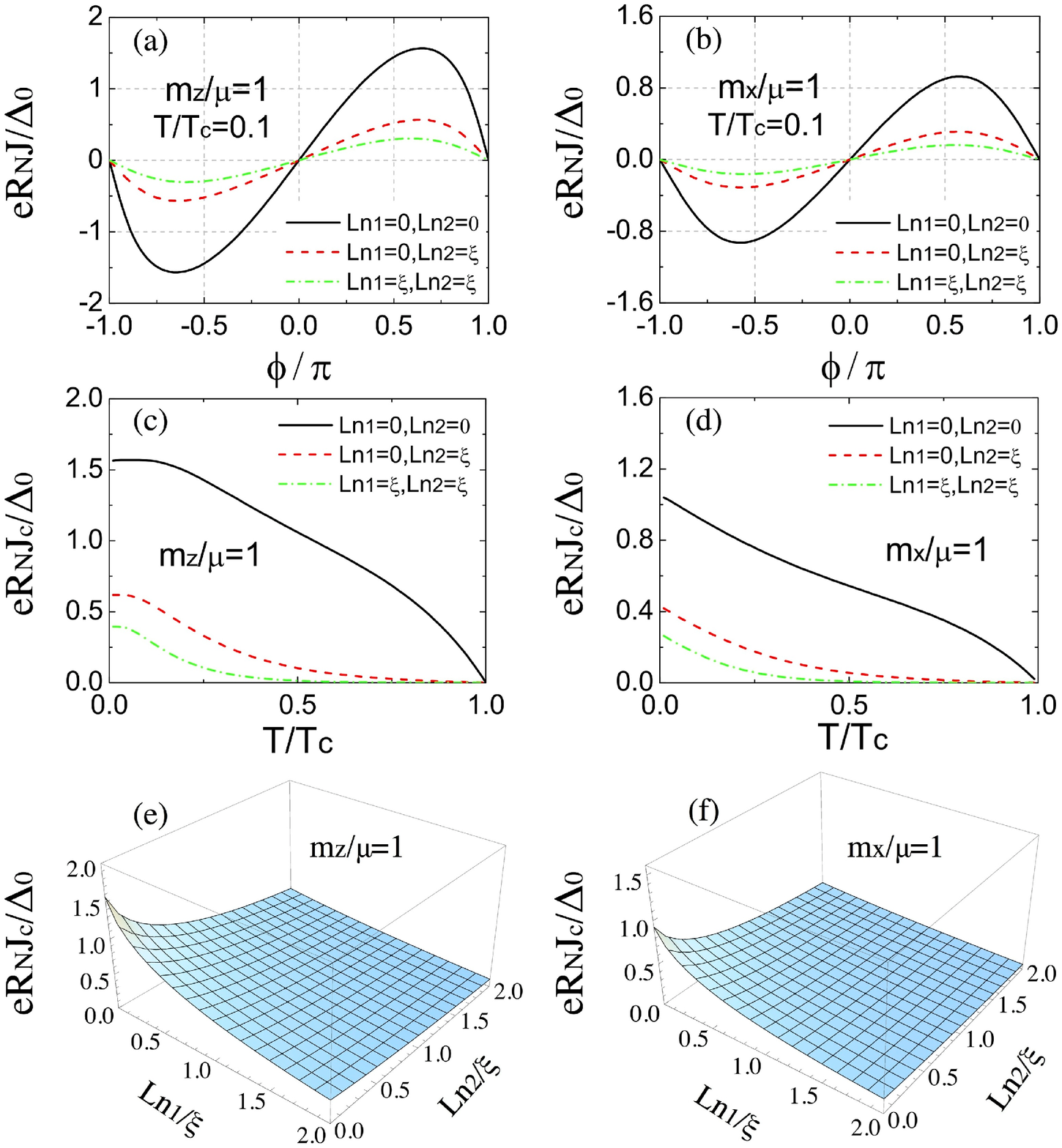}
\end{center}
\begin{center}
\caption{S/N/FI/N'/S junction: current-phase relation for 3 cases of the N layer length $L_{n1}$ and $L_{n2}$ for (a) $m_{z}/\mu=1$ and (b) $m_{x}/\mu=1$. (c)(d) Temperature dependence of critical current corresponding to (a) and (b), respectively. (e)(f) Critical Josephson current as a function of $L_{n1}$ and $L_{n2}$. The temperature is chosen as $T=0.1T_{c}$. Other parameters are chosen as the same as in Fig.\ref{fig5}. }
\label{fig6}
\end{center}
\end{figure}

We now consider S/N/FI/N'/S Josephson junctions. The length of N layer on
the two side of FI is denoted as $L_{n1}$ and $L_{n2}$. When $L_{n1}$ and $%
L_{n2}$ is on the superconducting coherence length scale, the junctions
become long-junctions. The influence of N layer between FI and S is shown in
Fig.\ref{fig6}. From Figs.\ref{fig6}(a) and (b), we can see that the
current-phase relation still retains the non-sinusoidal shape for different
values of $L_{n1}$ and $L_{n2}$ in low temperature limit. Throughout our
study, we have not found the sawtooth behavior of current-phase relation
involving magnetization in the long junction and low temperature limit, as
shown in Fig.\ref{fig5}(c). This is because the magnetization makes the
Andreev bound states gapped for most values of $k_{y}$ \cite{Tanaka2009}.
The derivative of energy dispersion which creates Josephson current will be
a smoother function of phase than that in S/N/S junction. For the
temperature dependence of the critical current, we can see that it behaves
qualitatively different in low temperature limit for $m_{z}$ and $m_{x}$ as
shown in Figs.\ref{fig6}(c) and (d), respectively. For $m_{z}$ case, the
critical current $J_{c}$ saturates at a constant value, which has been
revealed by the previous work \cite{Linderprb10}. However, for the $m_{x}$
case, it shows a Kulik-Ome'lyanchuk type of critical current \cite{KO} which
has linear low-temperature behavior. We interpret it as a result of the
enhanced zero-energy LDOSs for $m_{x}$ magnetization as illustrated in Figs.%
\ref{fig3}(b)(d)(f). In the high temperature limit, it is shown that for both $m_{z}$ and $m_{x}$ cases, $J_{c}$
is a concave function while it crosses over to a convex function with
increasing $L_{n1}$ (or $L_{n2}$). This behavior is similar to the S/N/S
junction. Figures.\ref{fig6}(e) and (f) represent the critical current as a
function of the length $L_{n1}$ and $L_{n2}$, for different direction of
magnetization.
It is worth noting that, although the interlayer N in the
S/N/FI/N' junction could generate resonant spikes in the transport
phenomena,
e.g., spikes in Figs.\ref{fig2} and \ref{fig3}, we find no
oscillatory behavior in either current-phase relation or critical current as
a function of length N (or N'). The critical current decreases monotonically
with the length $L_{n1}+L_{n2}$.


\section{IV. SUMMARY}
In summary, we theoretically studied the S/N/FI/N', S/N/S  and S/N/FI/N'/S junctions on the surface of 3D topological insulator.
We have constructed a formula to obtain Green's function.
The conductance spectra and local density of states in S/N/FI/N' junction
show resonant spikes due to the Andreev bound states.
The calculated current phase relation and temperature dependence of
critical current in the Josephson junctions are consistent with recent experiments in S/N/S junction.
We have also calculated current phase relation and temperature dependence of
critical current in S/N/FI/N'/S junction.
The non-sinusoidal current phase relation can be expected for short junctions.
We hope the obtained results will be confirmed by experiments in the near future.

\section{ACKNOWLEDGEMENTS} We thank V.V. Ryazanov, A.A. Golubov, Y. Asano  and M. Sato for
valuable discussions. This work
was supported in part by Grants-in-Aid for Scientific
Research from the Ministry of Education, Culture,
Sports, Science and Technology of Japan (Topological
Quantum Phenomena No.22103005 and No.25287085), by the German-Japanese research unit FOR1483 on "Topotronics", and by the
Ministry of Education and Science of the Russian Federation Grant No.14Y.26.31.0007.

\section{APPENDIX: WAVE FUNCTIONS}
The wave functions in the N interlayer are%
\begin{eqnarray}
\psi _{i}(x) &=&\sum\limits_{\lambda =1}^{4}s_{\lambda }^{i}\hat{N}_{\lambda
}e^{ik_{n,\lambda }x}, \\
\tilde{\psi}_{i}(x) &=&\sum\limits_{\lambda =1}^{4}\tilde{s}_{\lambda }^{i}%
\hat{N}_{\lambda }e^{ik_{n,\lambda }x},
\end{eqnarray}%
where
\begin{eqnarray}
\hat{N}_{1\left( 2\right) } &=&\left[ v_{f}\left( ik_{n,1\left( 2\right)
}+k_{y}\right) ,\mu +E,0,0\right] ^{T}, \\
\hat{N}_{3\left( 4\right) } &=&\left[ 0,0,v_{f}\left( ik_{n,3\left( 4\right)
}-k_{y}\right) ,-\mu +E\right] ^{T},
\end{eqnarray}%
with%
\begin{eqnarray}
k_{n,1\left( 2\right) } &=&\pm \sqrt{(\mu +E)^{2}/v_{f}^{2}-k_{y}^{2}}=\pm
k_{n}^{+}, \\
k_{n,3\left( 4\right) } &=&\pm \sqrt{(\mu -E)^{2}/v_{f}^{2}-k_{y}^{2}}=\pm
k_{n}^{-}.
\end{eqnarray}%
For the FI interlayer, we find that%
\begin{eqnarray}
\psi _{i}(x) &=&\sum_{\lambda =1}^{4}f_{\lambda }^{i}\hat{F}_{\lambda
}e^{ik_{\lambda }^{f}x}, \\
\tilde{\psi}_{i}(x) &=&\sum_{\lambda =1}^{4}\tilde{f}_{\lambda }^{i}\hat{F}%
_{\lambda }e^{ik_{\lambda }^{f}x},
\end{eqnarray}
where
\begin{subequations}
\begin{alignat}{4}
\hat{F}_{1}& =\left[ iv_{f}k_{1}^{f}+\left( v_{f}k_{y}+m_{x}\right)
,E-m_{z},0,0\right] ^{T}, & & & & & & \\
\hat{F}_{2}& =\left[ E+m_{z},-iv_{f}k_{2}^{f}+\left( v_{f}k_{y}+m_{x}\right)
,0,0\right] ^{T}, & & & & & & \\
\hat{F}_{3}& =\left[ 0,0,-iv_{f}k_{3}^{f}+\left( v_{f}k_{y}-m_{x}\right)
,E+m_{z}\right] ^{T}, & & & & & & \\
\hat{F}_{4}& =\left[ 0,0,E-m_{z},iv_{f}k_{4}^{f}+\left(
v_{f}k_{y}-m_{x}\right) \right] ^{T}, & & & & & &
\end{alignat}%
with
\end{subequations}
\begin{subequations}
\begin{alignat}{4}
& k_{1}^{f}=-\varsigma _{1}\sqrt{E^{2}-m_{z}^{2}-\left(
v_{f}k_{y}+m_{x}\right) ^{2}},\newline
& & & & & & \\
& k_{2}^{f}=\varsigma _{1}\sqrt{E^{2}-m_{z}^{2}-\left(
v_{f}k_{y}+m_{x}\right) ^{2}}, & & & & & & \\
& k_{3}^{f}=\varsigma _{2}\sqrt{E^{2}-m_{z}^{2}-\left(
v_{f}k_{y}-m_{x}\right) ^{2}},\newline
& & & & & & \\
& k_{4}^{f}=-\varsigma _{2}\sqrt{E^{2}-m_{z}^{2}-\left(
v_{f}k_{y}-m_{x}\right) ^{2}}, & & & & & &
\end{alignat}%
and $\varsigma _{1\left( 2\right) }=\mathrm{sgn}\left( v_{f}k_{y}\pm m_{x}\right) $.
The wave functions in the N' region of S/N/FI/N' are
\end{subequations}
\begin{subequations}
\begin{alignat}{4}
& \psi _{1}(x)=c_{1}\hat{C}_{1}e^{ik_{n}^{+}x}+d_{1}\hat{C}%
_{2}e^{-ik_{n}^{-}x},\newline
& & \\
& \psi _{2}(x)=c_{2}\hat{C}_{2}e^{-ik_{n}^{-}x}+d_{2}\hat{C}%
_{1}e^{ik_{n}^{+}x},\newline
& & \\
& \psi _{3}(x)=\hat{C}_{3}e^{-ik_{n}^{+}x}+a_{3}\hat{C}%
_{2}e^{-ik_{n}^{-}x}+b_{3}\hat{C}_{1}e^{ik_{n}^{+}x},\newline
& & \\
& \psi _{4}(x)=\hat{C}_{4}e^{ik_{n}^{-}x}+a_{4}\hat{C}%
_{1}e^{ik_{n}^{+}x}+b_{4}\hat{C}_{2}e^{-ik_{n}^{-}x}. & &
\end{alignat}%
and
\end{subequations}
\begin{subequations}
\begin{alignat}{4}
& \tilde{\psi}_{1}(x)=\tilde{c}_{1}\hat{D}_{1}e^{ik_{n}^{+}x^{\prime }}+%
\tilde{d}_{1}\hat{D}_{2}e^{-ik_{n}^{-}x^{\prime }},\newline
& & & & & & \\
& \tilde{\psi}_{2}(x)=\tilde{c}_{2}\hat{D}_{2}e^{-ik_{n}^{-}x^{\prime }}+%
\tilde{d}_{2}\hat{D}_{1}e^{ik_{n}^{+}x^{\prime }},\newline
& & & & & & \\
& \tilde{\psi}_{3}(x^{\prime })=\hat{D}_{3}e^{-ik_{n}^{+}x^{\prime }}+\tilde{%
a}_{3}\hat{D}_{2}e^{-ik_{n}^{-}x^{\prime }}+\tilde{b}_{3}\hat{D}%
_{1}e^{ik_{n}^{+}x^{\prime }},
& & & & & & \\
& \tilde{\psi}_{4}(x^{\prime })=\hat{D}_{4}e^{ik_{n}^{-}x^{\prime }}+\tilde{a%
}_{4}\hat{D}_{1}e^{ik_{n}^{+}x^{\prime }}+\tilde{b}_{4}\hat{D}%
_{2}e^{-ik_{n}^{-}x^{\prime }}. & & & & & &
\end{alignat}%
with
\end{subequations}
\begin{equation}
k_{n}^{\pm }\equiv \sqrt{%
(\mu \pm E)^{2}/v_{f}^{2}}\cos \theta _{n}^{\pm },
\end{equation}%
The spinors are given by
\begin{subequations}
\begin{alignat}{4}
& \hat{C}_{1}(\hat{D}_{3}) & =& [i,\pm e^{\pm i\theta _{n}^{+}},0,0]^{T}, & &
& & \\
& \hat{C}_{2}(\hat{D}_{4}) & =& [0,0,\pm 1,ie^{\pm i\theta _{n}^{-}}]^{T}, &
& & & \\
& \hat{C}_{3}(\hat{D}_{1}) & =& [ie^{\pm i\theta _{n}^{+}},\mp 1,0,0]^{T}, &
& & & \\
& \hat{C}_{4}(\hat{D}_{2}) & =& [0,0,\mp e^{\pm i\theta _{n}^{-}},i]^{T}. & &
& &
\end{alignat}

Also, the conductance can be given as
\end{subequations}
\begin{equation}
\sigma _{s}=\sigma _{0}\int dk_{y}\mathrm{Re}\left[ 1+\frac{k_{n}^{-}}{k_{n}^{+}}%
\left\vert a_{3}\right\vert ^{2}-\left\vert b_{3}\right\vert ^{2}\right]
\end{equation}%
where $\sigma _{0}$ is a constant parameter determined by the geometry of
junctions.

\end{document}